\documentclass[twocolumn,showpacs,preprintnumbers,amsmath,amssymb,prb,superscriptaddress,floatfix]{revtex4}
\usepackage{bm}
\usepackage{subfigure}
\usepackage{amssymb}
\usepackage{graphicx}
\usepackage{dcolumn}
\usepackage{bm}
\usepackage{epsfig}
\usepackage{color}

\begin{document}
\title{Optimal Condition for Strong Terahertz Radiation from Intrinsic Josephson Junctions}

\author{Feng Liu}
\affiliation{International Center for Materials Nanoarchitectonics (WPI-MANA), National Institute for
Materials Science, Tsukuba 305-0044, Japan}
\affiliation{Graduate School of Pure and Applied Sciences, University of Tsukuba, Tsukuba 305-8571, Japan}
\author{Shi-Zeng Lin}
\affiliation{International Center for Materials Nanoarchitectonics (WPI-MANA), National Institute for
Materials Science, Tsukuba 305-0044, Japan}
\author{Xiao Hu}
\affiliation{International Center for Materials Nanoarchitectonics (WPI-MANA), National Institute for
Materials Science, Tsukuba 305-0044, Japan}
\affiliation{Graduate School of Pure and Applied Sciences, University of Tsukuba, Tsukuba 305-8571, Japan}
\affiliation{Japan Science and Technology Agency, 4-1-8 Honcho, Kawaguchi, Saitama 332-0012, Japan}

\begin{abstract}
In order to enhance the radiation power in terahertz band based on the intrinsic Josephson junctions of Bi$_2$Sr$_2$CaCu$_2$O$_{8+\delta}$ single crystal, we investigate a long cylindrical sample embedded in a dielectric material. Tuning the dielectric constant, the radiation power has a maximum which is achieved when it equals the dissipation caused by Josephson plasma. This yields the optimal dielectric constant of wrapping material in terms of the properties of BSCCO single crystal. The maximal radiation power is found proportional to the product of the typical superconducting current squared and the typical normal resistance, or the gap energy squared divided by the typical normal resistance, which offers a guideline for choosing superconductor as a source of strong radiation. By introducing an anti-reflection layer, we can build a compact device with the BSCCO cylinder and two wrapping dielectric layers with finite thicknesses.
\end{abstract}
\date{May 29, 2012}
\maketitle

\section{Introduction}
Electromagnetic (EM) waves are vastly used in our daily life and their generation is considered as one of most developed scientific and technological fields. However there still exists a region around terahertz (THz) frequencies which have wide applications such as radar, drug detection, safety-check, and so on, lacking of compact solid-state generator.\cite{Zhang_2007,Tonouchi_2007}

It has been known for a long time that the Josephson junctions can work as oscillators to excite high-frequency EM waves.\cite{Hu_Lin_2010} Artificial Josephson junctions were used first,\cite{Dmitrenko_1965,dayem_1966,zimmerman_1966,pedersen_1976,finnegan_1972,Jain_1984,Heiden_1993,wiesenfeld_1994,Darula_1999,Barbara_1999}
but the frequency is below THz due to the small superconducting energy gap in conventional superconductors. The discovery of intrinsic Josephson junctions (IJJs) in layered high-T$_{\rm{c}}$ superconductor provides a fantastic chance to achieve strong coherent THz radiation.\cite{Kleiner_1992} The advantages of IJJs over conventional low-temperature junctions are as follows. First, the junctions are homogeneous at the atomic scale guaranteed by the high quality of single crystals, and second the superconductivity gap is large, which in principle permits the frequency to cover the whole range of THz band. Much effort has been made to stimulate powerful radiation based on IJJs. \cite{Hu_Lin_2010,sakai_1993,Tachiki_1994,Koyama_1995,Tachiki_1999,kume_1999,Kleiner_2000,Iguchi_2000,Mochiku_2000,Takahshi_2000,batov_2006,Bulaevskii_2006,Bulaevskii_JSNM_2006,Wang_2007,Bulaev_2008,Lin_Hu_Tachiki_2008}

An experimental breakthrough was made in 2007, where the THz radiation was observed based on IJJs from Bi$_2$Sr$_2$CaCu$_2$O$_{8+\delta}$ (BSCCO) mesa by biasing a dc voltage.\cite{Ozyuzer_2007} The key experimental results are following: the frequency of the radiation EM wave and the bias voltage obey the ac Josephson relation, and the frequency equals one of the cavity modes determined by the lateral size of mesa.\cite{Ozyuzer_2007}

A novel $\pi$ kink state has been proposed, which can explain the important experimental results.\cite{Lin_Hu_2008,Kosh_2008,Hu_Lin_PRB_2008,Lin_Hu_PRB_2009,Hu_Lin_2009} This state is characterized by static $\pm\pi$ phase kinks in the lateral directions of the mesa, which align themselves alternatingly along the $c$-axis. The $\pi$ phase kinks provide a strong coupling between the uniform dc current and the cavity modes, which permits large supercurrent flow into the system at the cavity resonances, thus enhances the plasma oscillation and radiates strong EM wave from the mesa edge.\cite{Hu_Lin_2010}

\begin{figure}[t]
\begin{center}
\leavevmode
\includegraphics[clip=true,width=0.85\columnwidth]{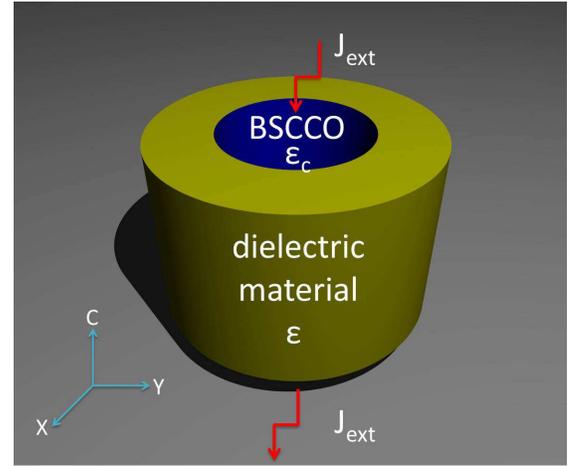}
\caption{(Color online) Schematic view of proposed device with a thick BSCCO cylinder embedded in a dielectric material.}
\end{center}
\end{figure}

Even with this novel mechanism, the radiation power based on IJJs is still weak due to the small radiation area limited by the mesa thickness ($\sim1\mu m$), and enhancement of the radiation power remains a problem. Here we propose to use a long cylindrical BSCCO single crystal embedded in a dielectric material as shown schematically in Fig.1. In this way, on one hand we can increase the radiation area, and on the other hand we can have a cavity. Then there is a question: what is the optimal dielectric material for strong radiation?

In order to answer this question, we solve the cavity modes and quality factor through Maxwell's equations, and analyze the phase dynamics in IJJs according to the coupled sine-Gordon equations. We find that the radiation power has a maximum when the dielectric constant is tuned. It turns out that the maximum is achieved when the radiation power equals the plasma dissipation. This gives us the optimal dielectric constant of wrapping material in terms of the property of the superconductive single crystal. The maximal radiation power is proportional to the typical superconducting current squared and the typical normal resistance, or the gap energy squared divided by the typical normal resistance.

The remaining part of the paper is organized as follows. In Sec. II, we discuss the cavity modes and quality factor of the system according to Maxwell's equations. In Sec. III, we analyze the phase dynamics associated with $\pi$ phase kink state in IJJs based on coupled sine-Gordon equations. In Sec. IV, we derive the optimal condition for strong radiation and the maximal radiation power explicitly. Sec. V is devoted to discuss the device with anti-reflection (AR) coating. Discussion and summary are given in Sec. VI and Sec. VII respectively.

\section{Cavity mode and Q factor}
\subsection{Cavity mode}
For simplicity we consider first a long BSCCO cylinder embedded in a dielectric material. The dielectric constant of wrapping material should be larger than that of BSCCO sample to ensure the existence of radiating modes. As we focus on the coherent radiation,  the EM field is considered uniform along the $c$ direction.

For an open (or a radiating) system, the EM waves should be described by complex numbers. By considering the Helmholtz wave equation and the continuity conditions of EM wave at the interface between BSCCO single crystal and wrapping material, the complex eigen wave number $k_{mn}$ can be given by\cite{Jackson_1998}
\small
\begin{eqnarray}
  \label{eigenvalue}
 \left({J_m' } H_m\right)^2+\varepsilon_r\left(J_m {H_m'}\right)^2+  {(1+\varepsilon_r)}{J_m'} J_m H_m {H_m'}=0,
\end{eqnarray}
\normalsize
 where $J_m(\sqrt{\varepsilon_c} k_{mn} R)$ and $H_m(\sqrt{\varepsilon}k_{mn} R)$ are the first kind of Bessel and Hankel functions with R the radius of BSCCO cylinder, $\varepsilon_r=\varepsilon/\varepsilon_c$ is the ratio between dielectric constants of wrapping material and BSCCO single crystal; ``prime''s denote the derivatives with respect to $R$.

 There are two kinds of cavity modes, namely perfect magnetic conductor (PMC) like one, and perfect electric conductor (PEC) like one.\cite{Jackson_1998} Here we take the PMC-like  (1,1) mode as an example, and the extension for other modes is straightforward. The spatial part of electric field in the $c$ direction for PMC-like (1,1) mode inside the BSCCO cylinder is given by\cite{Hu_Lin_2009}
\begin{eqnarray}
  \label{eigenmode}
  g_{11}(\mathbf{r})=J_1(\sqrt{\varepsilon_c}k_{11}\rho)\cos{\phi}
\end{eqnarray}
with the cylindrical coordinate $ \mathbf{r}=\left(\rho,\phi\right)$, $k_{11}$ the eigen wave number of  PMC-like (1,1) mode, as shown in Fig. 2.


\begin{figure}[t]
\begin{center}
\leavevmode
\includegraphics[clip=true,width=0.999\columnwidth]{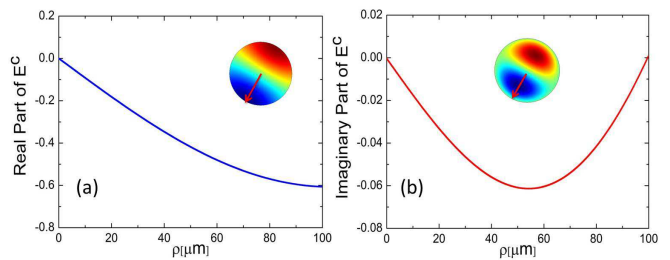}
\caption{(Color online) Distribution of (a) the real part and (b) the imaginary part of electric field in the $c$ direction for PMC-like (1,1) mode.}
\end{center}
\end{figure}

The angular eigen frequency, or equivalently the real part of eigen wave number $k_{11}$ in a dimensionless form, is given by
 \begin{equation}
   \omega_0=\dfrac{\chi_{11}c}{R\sqrt{\varepsilon_{c}}}
 \end{equation}
 with $\chi_{11}\approx 1.84$ standing for the first node of $J_1'$. For $R=100\mu m$ and $\varepsilon_{c}=16$, $f_0=\omega_0/2\pi=0.22\text{THz}$. The eigen frequency will not change with the dielectric constant of wrapping material as far as it is larger than the one of BSCCO sample.
\subsection{Quality factor}
We use quality factor
 \begin{eqnarray}
    \label{Q}
   Q\equiv \omega_0\dfrac{\rm{Energy \quad Stored}}{\rm{Radiation \quad Power}},
 \end{eqnarray}
 to characterize the energy loss caused by the radiation in the open system.\cite{Jackson_1998} Considering that a cavity mode oscillates harmonically with time, the quality factor can be rewritten as
   \begin{eqnarray}
    \label{Q2}
    Q=\dfrac{k_{\rm{r}}}{2k_{\rm{i}}},
 \end{eqnarray}
 where $k_{\rm{r}}$ and $k_{\rm{i}}$ are the real and imaginary parts of wave number.
 The $\varepsilon_r$ dependence of the quality factor can be evaluated from Eqs.(\ref{eigenvalue}) and (\ref{Q2}). As displayed in Fig.3, the quality factor increases with $\varepsilon_{\rm{r}}$ approximately in the form:
\begin{eqnarray}
    Q\approx 0.66 \sqrt{\varepsilon_r}-0.13.
    \label{Q3}
\end{eqnarray}
\begin{figure}[htb]
\label{Qfactor}
\begin{center}
\leavevmode
\includegraphics[clip=true,width=0.9\columnwidth]{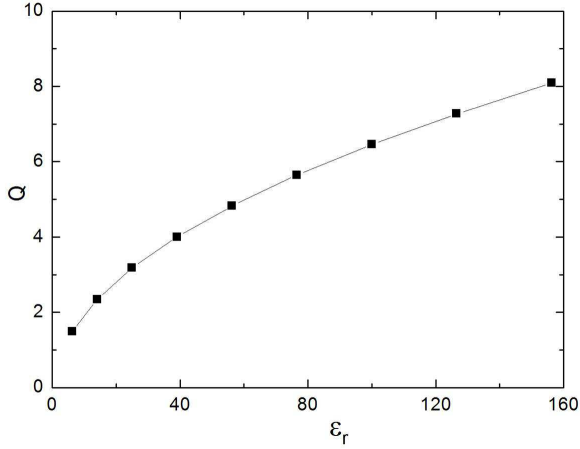}
\caption{Dependence of the quality factor on the ratio of dielectric constants between wrapping material and BSCCO single crystal for PMC-like (1,1) mode.}
\end{center}
\end{figure}

\section{phase dynamics in IJJs}
The phase dynamics in IJJs under bias voltage can be described by the inductively coupled sine-Gordon equations, which are given in the dimensionless form as \cite{Lin_Hu_2008,Kosh_2008,Hu_Lin_PRB_2008,Lin_Hu_PRB_2009,Hu_Lin_2009,Hu_Lin_2010}
\begin{eqnarray}
  \bigtriangleup \gamma_l=\left(1-\zeta\bigtriangleup^{(2)}\right)\left
  (\sin{\gamma_l}+\beta\partial_t \gamma_l+\partial^2_t \gamma_l-J_{\rm{ext}}\right),
  \label{sineGodon}
\end{eqnarray}
where the lateral length is scaled by the penetration depth $\lambda_{c}$ and time is scaled by the inverse of intrinsic plasma frequency $\lambda_{c} \sqrt{\varepsilon_{c}}/c$; $\gamma_l$ is the gauge-invariant phase difference at the $l$th junction, $\zeta\equiv(\lambda_{ab}/s)^2 $ is the inductive coupling which is in order of $10^5$ ($\lambda_{ab}$ the penetration depth of lateral plane and $s$ standing for the period of BSCCO lattice in the c direction), $\beta \equiv 4\pi \sigma_{c} \lambda_{c}/ c\sqrt{\varepsilon_{c}}$ is the normalized c-axis conductivity of BSCCO sample, $\bigtriangleup$ is the Laplace operator in lateral directions,
$\bigtriangleup^{(2)}\gamma_l=\gamma_{l+1}+\gamma_{l-1}-2\gamma_l$ is the second-order difference operator along the $c$-axis.

The general form of $\pi$ phase kink solution\cite{Lin_Hu_2008,Kosh_2008,Hu_Lin_PRB_2008,Lin_Hu_PRB_2009,Hu_Lin_2009,Hu_Lin_2010} for the coupled sine-Gordon equations is given by
\begin{eqnarray}
  \gamma_l(\mathbf{r},t)=\omega t+Ag(\mathbf{r})e^{i\omega t}+(-1)^l\gamma^s(\mathbf{r}),
  \label{pikink}
\end{eqnarray}
where the first term is the rotating phase accounting for the finite dc bias voltage; the second term stands for the cavity term of plasma oscillation, where $A$ is the complex amplitude, $g(\mathbf{r})$ is the cavity mode of electric field in the $c$ direction which is complex for the present open system as we solved above, and the frequency $\omega$ is given by the bias voltage based on the ac Josephson relation; the third term is the $\pi$ phase kink which carries the strong interjunction coupling via the $l$ dependence.


Inserting Eq.(\ref{pikink}) into Eq.(\ref{sineGodon}) and omitting higher harmonics for $|A|\ll 1$ we arrive at
the following coupled equations
\begin{eqnarray}
     J_{\rm{ext}}=\beta\omega+\dfrac{\int \rm{Re}{(Ag)}\cos{\gamma^s}d\Omega}{2\int d\Omega},
    \label{external current}
\end{eqnarray}
\begin{eqnarray}
 A=\dfrac{\int g^* \cos{\gamma^s}d\Omega}{2i(\omega^2-k^2-i\omega\beta )\int  \lvert g \rvert^2 d\Omega},
 \label{A}
 \end{eqnarray}
\begin{eqnarray}
\bigtriangleup \gamma^s=2\zeta \rm{Im}(Ag)\sin{\gamma^s},
\label{pi}
\end{eqnarray}
with the integration over the lateral area of BSCCO cylinder. From Eq.(\ref{pi}) it is clear that the $\pi$ phase kink arises at the node of cavity electric field.\cite{Lin_Hu_2008,Hu_Lin_PRB_2008,Lin_Hu_PRB_2009,Hu_Lin_2009,Hu_Lin_2010}

 The radiation power and plasma dissipation per unit length in the $c$ direction are given respectively by
 \begin{eqnarray}
 P_{\rm{r}}=\dfrac{\omega}{2Q}\int{\lvert E \rvert ^2} d\Omega,\quad P_{\rm{dp}}=\dfrac{\beta}{2}\int{\lvert E \rvert ^2} d\Omega,
 \label{radiatonpower}
 \end{eqnarray}
 where $E=Ag$ is the spatial part of electric field in the $c$ direction.

 For a given $Q$ factor, or equivalently a given dielectric constant (see Eq.(\ref{Q3})), the voltage dependence of current injection and radiation power can be evaluated by solving Eqs.(\ref{external current}) to (\ref{radiatonpower}). The result for $Q=73.73$ (or $\varepsilon=2.01 \times 10^5$) is displayed in Fig. 4.
 \begin{figure}[t]
\begin{center}
\leavevmode
\includegraphics[clip=true,width=0.9\columnwidth]{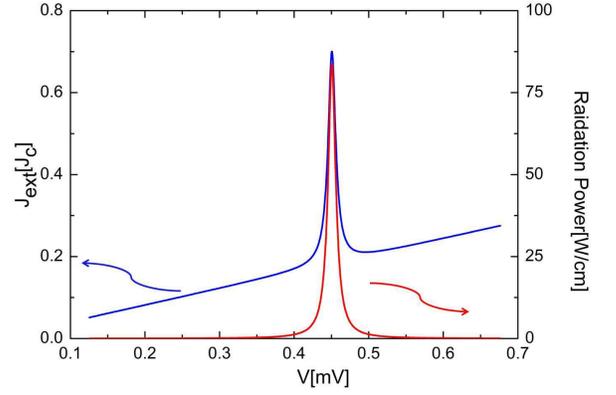}
\caption{(Color online) Dependence of  current injection and radiation power on the  bias voltage for $Q$=73.72 and $\beta$ =0.05. The resonance takes place at V=0.45mV, which corresponds to the eigen frequency 0.22 THz of the PMC-like (1,1) mode for R=100$\mu$m. }
\end{center}
\end{figure}
 \section{optimal radiation power}
 Now we investigate the $Q$-dependence of the radiation power and the dissipation caused by plasma oscillation.
 \begin{figure}[htb]
\label{OptimalPower}
\begin{center}
\leavevmode
\includegraphics[clip=true,width=0.9\columnwidth]{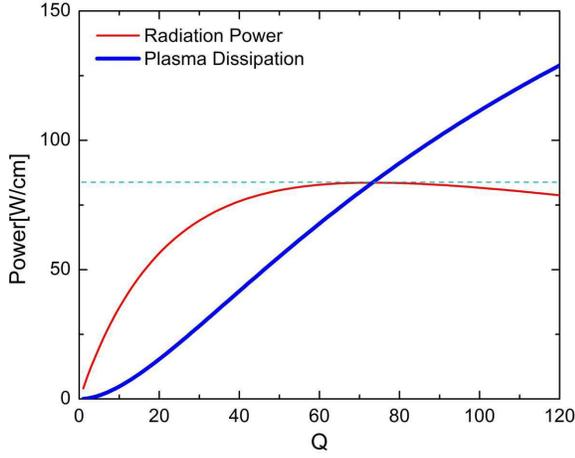}
\caption{(Color online) $Q$-factor dependence of radiation power and plasma dissipation  at resonance for the PMC-like (1,1) mode with $\beta=0.05$ and $R=100\mu m$. The radiation power reaches its maximum when it equals plasma dissipation.}
\end{center}
\end{figure}
As shown in Fig. 5, the radiation power reaches its maximum when it equals the plasma dissipation. The reason for the existence of a maximal radiation power is that there exist two competing factors in the system: a good cavity associated with large $Q$ makes a strong resonance on one hand, but it suppresses energy leak on the other hand.

In order to understand the optimal condition for radiation power, we check the relations between radiation power and plasma dissipation. The ratio between them is given by
\begin{eqnarray}
  \dfrac{P_{\rm{r}}}{P_{\rm{dp}}}=\dfrac{\omega}{Q\beta}.
  \label{ratio}
\end{eqnarray}
From the energy conservation we have
\begin{eqnarray}
  P_{\rm{r}}+P_{\rm{dp}}=\dfrac{S}{2}\omega(J_{\rm{ext}}-\omega \beta),
  \label{energy}
\end{eqnarray}
where $S$ is the lateral area of BSCCO cylinder and $\omega\beta$ is the normal ohmic current.

 From Eqs.(\ref{external current}) and (\ref{A}) the supercurrent density is given by
\begin{eqnarray}
  J_{\rm{ext}}-\beta \omega= \dfrac{\nu}{S} \rm{Re}\left[\dfrac{1}{i(\omega^2-k^2-i\omega\beta)}\right],
  \label{supercurrent}
\end{eqnarray}
with
\begin{eqnarray}
  \nu=\dfrac{\left(\int g_{\rm{r}}\cos{\gamma^s}d\Omega\right)^2 +\left(\int  g_{\rm{i}}\cos{\gamma^s}d\Omega\right)^2}{4\int \lvert g \rvert^2d\Omega},
\end{eqnarray}
where $g_{\rm{r}}$ and $g_{\rm{i}}$ are the real and imaginary parts of cavity mode. From Eqs.(\ref{ratio}) to (\ref{supercurrent}), the radiation power  at resonance where $\omega=k_{\rm{r}}=\omega_0$ can be given by
\begin{eqnarray}
  P_{\rm{r}}=\dfrac{\nu\omega_0/Q}{2(\omega_0/Q+\beta)^2}.
  \label{Optimal Pr}
\end{eqnarray}
It is then clear that $P_{\rm{r}}$ reaches the maximal value at
\begin{eqnarray}
    \label{Optimal Condition}
    Q=\dfrac{\omega_0}{\beta},
\end{eqnarray}
 where $P_{\rm{r}}=P_{\rm{dp}}$ from Eq.(\ref{ratio}).

 By substituting Eq.(\ref{Q3}) into Eq.(\ref{Optimal Condition}), the optimal dielectric constant of wrapping material for strong radiation can be obtained
 \begin{eqnarray}
  \dfrac{\varepsilon}{\varepsilon_{c}}\approx \left( \dfrac{\chi_{11}c\sqrt{\varepsilon_{c}}}{2.64\pi R\sigma_{c}}+0.20\right)^2.
\end{eqnarray}
 In Gaussian units, Eqs.(\ref{Optimal Pr}) and (\ref{Optimal Condition}) give
 \begin{eqnarray}
 P_{\rm{r}}=\frac{\nu}{8}I_{\rm{c}}^2 R_{c},
  \label{OptimalP}
\end{eqnarray}
where $I_{\rm{c}}\equiv J_{\rm{c}}\lambda_{c}^2=c\phi_0/8\pi^2s$ is the typical superconducting current with $\phi_0$ the magnetic flux quantum, $R_{c}\equiv1/\sigma_{c}\lambda_{c}^2$ is the typical $c$-axis resistance per unit length of BSCCO sample. This expression looks like the Ohm's law but we should notice that $I_{\rm{c}}$ is the typical superconducting current.

The Eq.(\ref{OptimalP}) can also be written as:
\begin{eqnarray}
 P_{\rm{r}}=\dfrac{\nu\pi^2\Delta^2}{32R_{\rm{c}}s^2e^2},
  \label{OptimalP2}
\end{eqnarray}
where $\Delta=ec\phi_0/4\pi^3 \lambda_{c}^2\sigma_{c}$ is the superconducting gap energy.

For the PMC-like (1,1) mode $\nu\approx 0.15$. Taking $I_{\rm{c}}=0.22\rm{A}$, $R_{c}=0.95\times10^5\Omega/\rm{cm}$, typical values for BSCCO,\cite{Ozyuzer_2007} the maximal radiation power is evaluated as $85\rm{W/cm}$.

\section{Setup with anti-reflection coating}
 In the above discussion we focus on the simple case in which the thickness of wrapping material is infinite, where the only one reflection of EM wave at the interface between BSCCO and the dielectric material determines the cavity frequency. In what following, we show that two wrapping layers with finite thicknesses can work in the same way (Fig. 6), when the dielectric constant and thickness of the outside layer are tuned appropriately such that the two reflections at the interfaces between the two wrapping layers and between the outer layer and vacuum cancel each other totally.
 \begin{figure}[b]
\centering
\includegraphics[clip=true,width=0.9\columnwidth]{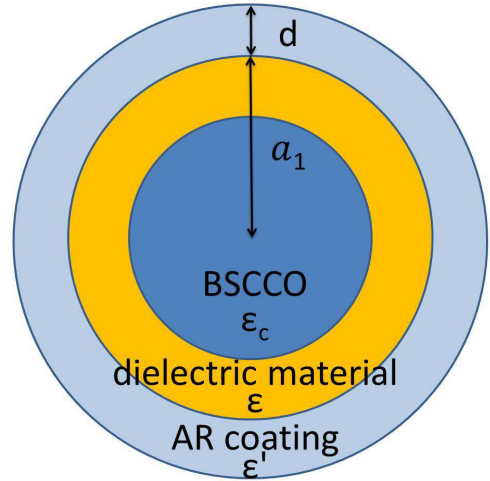}
\caption{(Color online) Schematic view of the proposed device with anti-reflection coating.}
\end{figure}
 The ratio between ingoing and outgoing of electric fields at the interface between two dielectric layers is given by
 \footnotesize
 \begin{widetext}
 \begin{eqnarray}
 \label{refectance}
   r=\frac{M_1{}_1H_m^{(1)}(k_{\rm{L}}a_1)H_m^{(1)}{}'(k_{\rm{R}}a_2)+M_1{}_2H_m^{(1)}{}'(k_{\rm{L}}a_1)H^{(1)}{}'(k_{\rm{R}}a_2)
-M_{21}H_m^{(1)}(k_{\rm{L}}a_1)H_m^{(1)}{}(k_{\rm{R}}a_2)-M_{22}H_m^{(1)}{}'(k_{\rm{L}}a_1)H_m^{(1)}{}(k_{\rm{R}}a_2)}
   {M_{21}H_m^{(2)}(k_{\rm{L}}a_1)H_m^{(1)}(k_{\rm{R}}a_2)+M_{22}H_m^{(2)}{}'(k_{\rm{L}}a_1)H_m^{(1)}{}(k_{\rm{R}}a_2)
   -M_{11}H_m^{(2)}(k_{\rm{L}}a_1)H_m^{(1)}{}'(k_{\rm{R}}a_2)-M_{12}H_m^{(2)}{}'(k_{\rm{L}}a_1)H_m^{(1)}{}'(k_{\rm{R}}a_2)},
 \end{eqnarray}
\normalsize
 with the transfer matrix
 \begin{eqnarray}
  \mathbf{M}=\begin{pmatrix}
    H^{(1)}_m(k_Ma_2) \quad -H^{(2)}_m(k_Ma_2)
    \\
    H_m^{(1)}{}'(k_Ma_2) \quad -H_m^{(2)}{}'(k_Ma_2)
    \end{pmatrix}
    \begin{pmatrix}
    H^{(1)}_m(k_Ma_1) \quad -H^{(2)}_m(k_Ma_1)
    \\
    H_m^{(1)}{}'(k_Ma_1) \quad -H_m^{(2)}{}'(k_Ma_1)
    \end{pmatrix}^{-1},
\end{eqnarray}
\end{widetext}
\normalsize
 where $k_{\rm{L}}$, $k_{\rm{M}}$ and $k_{\rm{R}}$ are the wave numbers in the inner wrapping layer, the AR coating and vacuum respectively, and $a_2=a_1+d$ (for definition see Fig. 6); ``prime''s denote the derivatives with respect to $a_1$ or $a_2$. Solving the equation $r=0$  the dielectric constant and thickness of AR coating can be determined as functions of $\varepsilon$ and $a_1$. Numerical results for typical parameters are displayed in Fig. 7.
\begin{figure}[b]
\centering
\leavevmode
\begin{minipage}[b]{0.9\columnwidth}
\includegraphics[clip=true,width=0.9\columnwidth]{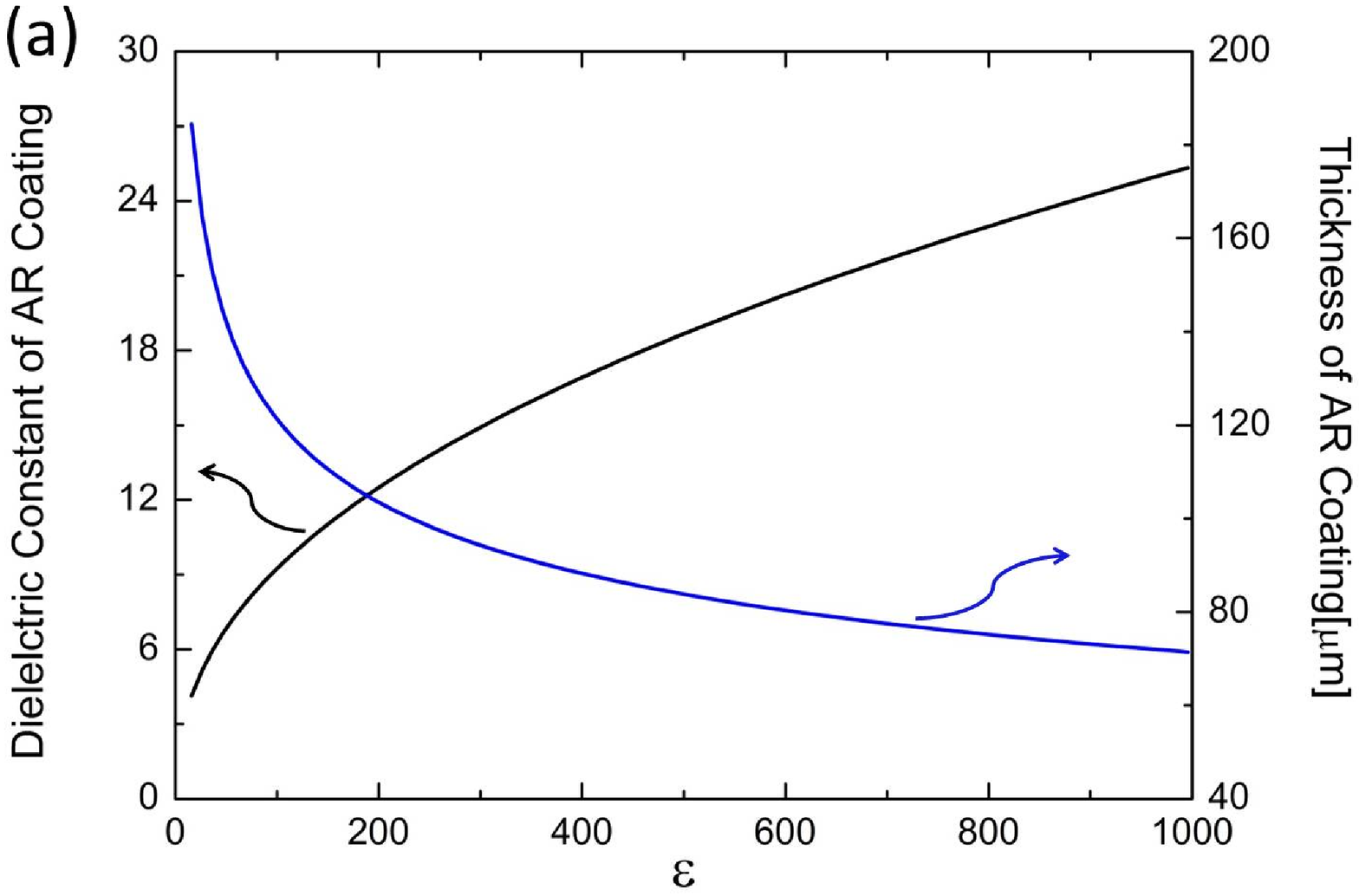}
\includegraphics[clip=true,width=0.9\columnwidth]{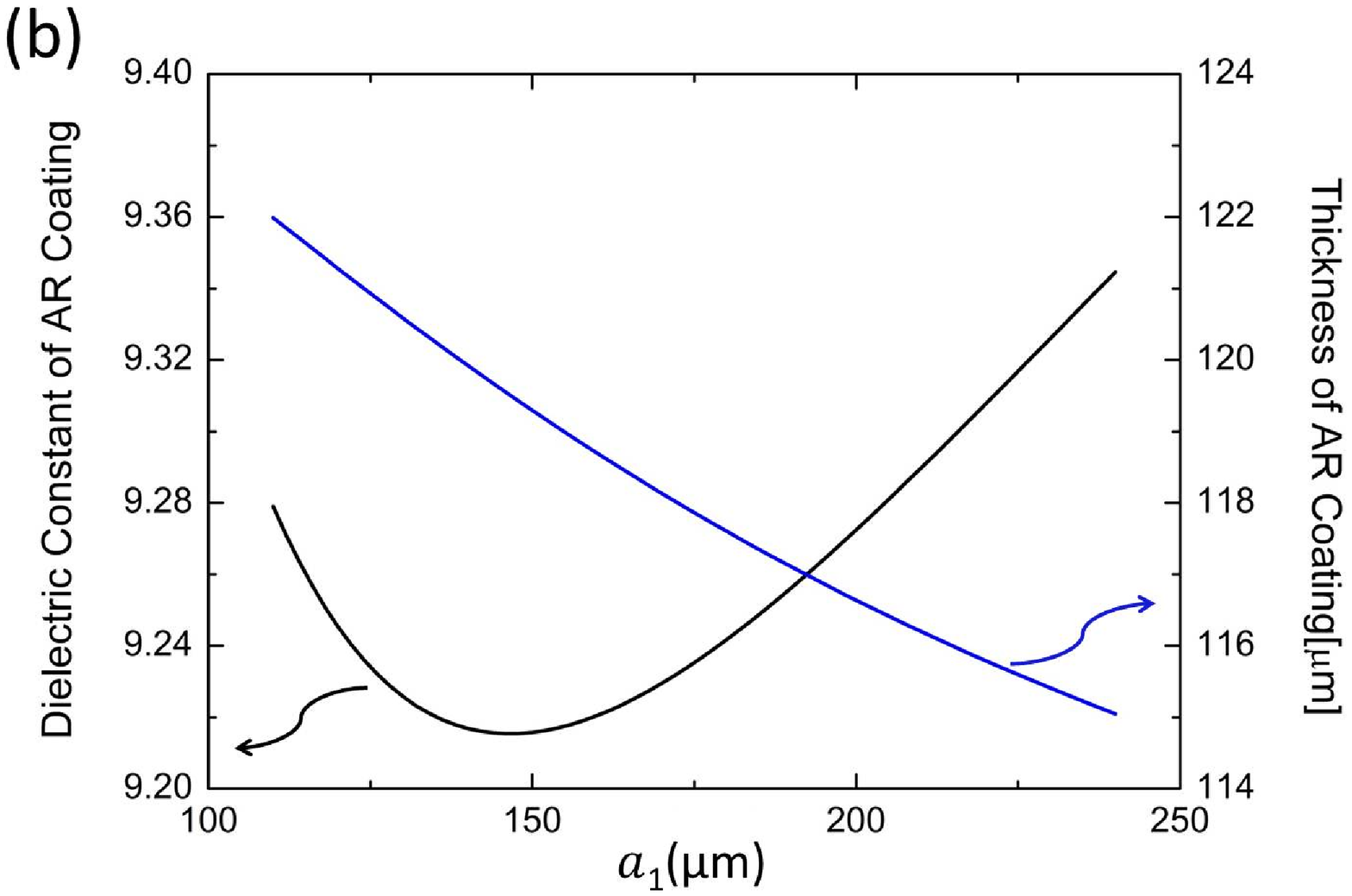}
\end{minipage}
\caption{(Color online) Dependence of  dielectric constant and thickness of anti-reflection coating on (a) the dielectric constant of the inner wrapping layer with $a_1=120\mu \rm{m}$, (b) the thickness of the inner wrapping layer with $\varepsilon=100$ of PMC-like (1,1) mode with eigen frequency $0.22$THz.
}
\end{figure}

\section{Discussions}
The maximal radiation power is not affected by the radius of the BSCCO cylinder. The reason for this is that when the radius of the BSCCO cylinder increases, more current can be  injected in, but the eigen frequency will decrease (proportional to the inverse of radius), with these two factors canceling each other. The optimal radiation power Eq.(\ref{OptimalP}) is apparently similar to the result in Ref. \onlinecite{wiesenfeld_1994}, but there the discussion  was limited to nonresonant junctions.

The result that the radiation power is maximal when it equals the dissipation can be considered as a case of Jacobi's law. In a recent work,\cite{Kras_2010} Krasnov reached numerically at a similar result for radiation in finite magnetic fields. This result does not depend on the geometry of BSCCO single crystal which can be seen from Eq.(17).

In the present work, we focus on the case that the Josephson plasma is uniform along a long crystal as established by the cavity formed by the dielectric material. The maximal radiation power is therefore given by power per unit length. Bulaevskii and Koshelev discussed the case that synchronization is achieved by the coupling between Josephson plasma and EM wave.\cite{Bula_Kosh_2007} Therefore, their maximal radiation power is achieved when the length of BSCCO single crystal at $c$-axis equals the EM wave length.

\section{Summary}
  To summarize, in order to enhance the radiation power in THz band we embed a long BSCCO cylinder in a dielectric material to increase the radiation area and meanwhile form a cavity. We find that the radiation power has a maximum when the dielectric constant is tuned. The maximum is achieved when the radiation power equals the plasma dissipation which yields the optimal dielectric constant of wrapping material in terms of the properties of the superconductive single crystal. The maximal radiation is proportional to the typical superconducting current squared and the typical normal resistance, or the gap energy squared divided by the typical normal resistance, which offers a guideline for choosing superconductor as a source of strong radiation. Adding an AR coating, we can build a compact device with the BSCCO cylinder and two wrapping dielectric layers with finite thicknesses.
\begin{acknowledgements}
  The authors want to thank Z. Wang and Q. F. Liang for useful discussions.
  This work was supported by WPI initiative on Materials Nanoarchitectonics, MEXT of Japan and by CREST, JST.
\end{acknowledgements}
\appendix*
\section{ANTI-REFLECTION COATING FOR CYLINDRICAL WAVE}

 For cylindrical propagation wave the electric field in the $c$ direction can be written by
 \begin{eqnarray}
   E_{c}(k,\rho)=A_1 H_m^{(1)}(k\rho)-A_2 H_m^{(2)}(k \rho),
 \end{eqnarray}
  where $H_m^{(1)}$ and $H_m^{(2)}$ are the first and the second kind of Hankel function, presenting the outgoing and ingoing cylindrical waves.  From Maxwell's equations, the magnetic field in the angular direction is given by
  \begin{eqnarray}
    B_\theta(k,\rho)=\dfrac{1}{i\omega}\left[A_1 H_m^{(1)}{}' (k \rho)-A_2 H_m^{(2)}{}' (k \rho)\right]
  \end{eqnarray}
 with ``prime'' denoting the derivative with respect to $\rho$.
 Considering the continuity of $E_{c}$ and $B_\theta$ at the interfaces between the inner wrapping layer and the AR coating and between the AR coating and vacuum (Fig. 6), we obtain
   \begin{widetext}
    \begin{eqnarray}
    \begin{pmatrix}
    H^{(1)}_m(k_{\rm{L}}a_1) \quad -H^{(2)}_m(k_{\rm{L}}a_1)
    \\
    H_m^{(1)}{}'(k_{\rm{L}}a_1) \quad -H_m^{(2)}{}'(k_{\rm{L}}a_1)
    \end{pmatrix}
    \begin{pmatrix}
    A_{{\rm{L}}1}
    \\
    A_{{\rm{L}}2}
    \end{pmatrix}
    =\begin{pmatrix}
    H^{(1)}_m(k_{\rm{M}}a_1) \quad -H^{(2)}_m(k_{\rm{M}}a_1)
    \\
    H_m^{(1)}{}'(k_{\rm{M}}a_1) \quad -H_m^{(2)}{}'(k_{\rm{M}}a_1)
    \end{pmatrix}
    \begin{pmatrix}
    A_{{\rm{M}}1}
    \\
    A_{{\rm{M}}{2}}
    \end{pmatrix},
    \\
     \begin{pmatrix}
    H^{(1)}_m(k_{\rm{M}}a_2) \quad -H^{(2)}_m(k_{\rm{M}}a_2)
    \\
    H_m^{(1)}{}'(k_{\rm{M}}a_2) \quad -H_m^{(2)}{}'(k_{\rm{M}}a_2)
    \end{pmatrix}
    \begin{pmatrix}
    A_{{\rm{M}}1}
    \\
    A_{{\rm{M}}2}
    \end{pmatrix}
    =\begin{pmatrix}
    H_m^{(1)}(k_{\rm{R}}a_2)
    \\
    H_m^{(1)}{}'(k_{\rm{R}}a_2)
    \end{pmatrix}
    A_{{\rm{R}}1},
    \label{transfer equation}
    \end{eqnarray}
    \end{widetext}
    \normalsize
 where $k_{\rm{L}}$, $k_{\rm{M}}$, and $k_{\rm{R}}$ are the wave vectors, $A_{{\rm{L}}1}$, $A_{{\rm{L}}2}$, and $ A_{{\rm{M}}1}$, $A_{{\rm{M}}2}$, and $A_{{\rm{R}}1}$ are the amplitudes of electric field in the inner wrapping layer, the AR coating and the vacuum respectively. Eliminating $A_{\rm{M}1}$ and $A_{\rm{M}2}$ in Eqs.(A.3) and (A.4) we have
 \begin{widetext}
 \begin{eqnarray}
    \mathbf{M}
    \begin{pmatrix}
    H^{(1)}_m(k_{\rm{L}}a_1) \quad -H^{(2)}_m(k_{\rm{L}}a_1)
    \\
    H_m^{(1)}{}'(k_{\rm{L}}a_1) \quad -H_m^{(2)}{}'(k_{\rm{L}}a_1)
    \end{pmatrix}
    \begin{pmatrix}
    A_{\rm{L}1}
    \\
    A_{\rm{L}2}
    \end{pmatrix}
    =
    \begin{pmatrix}
    H_m^{(1)}(k_{\rm{R}}a_2)
    \\
    H_m^{(1)}{}'(k_{\rm{R}}a_2)
    \end{pmatrix}
    A_{\rm{R}1}.
    \end{eqnarray}
    \end{widetext}
    \normalsize
Then we obtain the reflectance $r\equiv A_{{\rm{L}}2}/A_{{\rm{L}}1}$ as given in Eq.(\ref{refectance}) and the transmittance  $t\equiv A_{{\rm{R}}1}/A_{{\rm{L}}1}$ as
\footnotesize
\begin{widetext}
\begin{eqnarray}
      t=\frac{4i/\pi a_2}
   {M_{21}H_m^{(2)}(k_{\rm{L}}a_1)H_m^{(1)}(k_{\rm{R}}a_2)+M_{22}H_m^{(2)}{}'(k_{\rm{L}}a_1)H_m^{(1)}{}(k_{\rm{R}}a_2)
   -M_{11}H_m^{(2)}(k_{\rm{L}}a_1)H_m^{(1)}{}'(k_{\rm{R}}a_2)-M_{12}H_m^{(2)}{}'(k_{\rm{L}}a_1)H_m^{(1)}{}'(k_{\rm{R}}a_2)},
 \end{eqnarray}
 \end{widetext}
\normalsize
where the Wronskian formula
\begin{eqnarray}
 H_m^{(1)}(x)H_m^{(2)}{}'(x)-H_m^{(1)}{}'(x)H_m^{(2)}(x)=\dfrac{-4i}{\pi x},
\end{eqnarray}
\normalsize
has been used in order to obtain Eq.(A.7).

\end{document}